\documentclass[twocolumn,secnumarabic,showpacs,amssymb, amsmath,tightenlines,aps,prb]{revtex4}
\usepackage{longtable}
\usepackage{color}
\usepackage{hhline}
\usepackage{bm}
\usepackage{epsfig}
\usepackage{graphicx}

\begin{document}

\title[Interplay of magnetism, Fermi surface reconstructions, and hidden-order in the heavy-fermion material URu$_2$Si$_2$]
{Interplay of magnetism, Fermi surface reconstructions, and hidden-order in the heavy-fermion material URu$_2$Si$_2$}
\author{G.W. Scheerer$^{1}$, W. Knafo$^{1}$, D. Aoki$^{2}$, G. Ballon$^{1}$, A. Mari$^{3}$, D. Vignolles$^{1}$, and J. Flouquet$^{2}$}

\address{$^{1}$ Laboratoire National des Champs Magn\'{e}tiques Intenses, UPR 3228, CNRS-UJF-UPS-INSA, 143 Avenue de Rangueil,
31400 Toulouse, France.}
\address{$^{2}$ Institut Nanosciences et Cryog\'{e}nie, SPSMS, CEA-Grenoble, 17 rue des Martyrs, 38054 Grenoble, France.}
\address{$^{3}$ Laboratoire de Chimie de Coordination, 205, route de Narbonne, 31077  Toulouse cedex 4, France.}

\begin{abstract}

URu$_2$Si$_2$ is surely one of the most mysterious of the heavy-fermion compounds. Despite more than twenty years of experimental and theoretical works, the order parameter of the transition at $T_0 = 17.5$~K is still unknown. The state below $T_0$ remains called "hidden-order phase" and the stakes are still to identify the energy scales driving the system to this phase. We present new magnetoresistivity and magnetization measurements performed on very-high-quality single crystals in pulsed magnetic fields up to 60~T. We show that the transition to the hidden-order state in URu$_2$Si$_2$ is initially driven by a high-temperature crossover at around 40-50~K, which is a fingerprint of inter-site electronic correlations. In a magnetic field $\mathbf{H}$ applied along the easy-axis $\bf{c}$, the vanishing of this high-temperature scale precedes the polarization of the magnetic moments, as well as it drives the destabilization of the hidden-order phase. Strongly impurity-dependent magnetoresistivity confirms that the Fermi surface is reconstructed below $T_0$ and is strongly modified in a high magnetic field applied along $\mathbf{c}$, i.e. at a sufficiently-high magnetic polarization. The possibility of a sharp crossover in the hidden-order state controlled by a field-induced change of the Fermi surface is pointed out.

\end{abstract}


\maketitle

\section{Introduction}

Heavy-fermion physics is governed by the Kondo effect, which is a hybridization of $f$- and conduction electrons due to the closeness of the $f$-energy level to the Fermi energy. \cite{hewson93,flouquet05,lohneysen07} In Ce-based systems, $f$-electrons magnetic properties are driven by RKKY interactions and are very sensitive to pressure and chemical doping, which permit to tune a quantum phase transition between a paramagnetic regime and a (generally antiferro-) magnetic state. \cite{doniach77} In particular for the paramagnetic regime, a Fermi liquid picture is often adequate at low temperature and strongly-renormalized effective masses are related to intense magnetic fluctuations. For Ce$^{3+}$ and Yb$^{3+}$ Kramers ions, the crystal-field ground state is either a magnetic doublet or a quartet. Valence fluctuations occur between the magnetic trivalent state and empty Ce$^{4+}$ or fully-occupied Yb$^{2+}$ 4$f$ shells, these fluctuations being stronger when the Kondo temperature is bigger. In U-based compounds, valence fluctuations between the U$^{3+}$ (5$f^3$) and U$^{4+}$ (5$f^2$) configurations, which both have a large angular momentum, are reported. \cite{cooper79} When a renormalization to the U$^{4+}$ state  (5$f^2$ configuration) is appropriate, exotic properties can occur due to the possibility to also form a singlet ground state through the action of the crystal field. This can favor multipolar coupling as in Pr$^{3+}$-based systems, \cite{kuramoto09} also in $f^2$ configuration. URu$_2$Si$_2$ occupies a particular place in the heavy-fermion family \cite{mydosh11}: a second-order phase transition at the temperature $T_0 = 17.5$~K is reported by many experimental probes but, despite numerous propositions, no order parameter has been consensually associated to the phase below $T_0$, which is called "hidden-order" phase. The magnetic properties of URu$_2$Si$_2$ are that of a paramagnet in a mixed-valent state with strong inter-site correlations and presumably damped crystal-field effects. Enhanced magnetic fluctuations have been reported by inelastic neutron scattering at the wavevectors $\mathbf{Q}_1=(1.4,0,0)$ and $\mathbf{Q}_0=(1,0,0)$ [\onlinecite{broholm91}]. Under pressure, antiferromagnetic long-range ordering is stabilized above 0.5~GPa within the wavevector $\mathbf{Q}_0$, the ordered moments reaching 0.4~$\mu_B$/U at 1~GPa [\onlinecite{amitsuka07}]. A magnetic field along the easy magnetic axis $\mathbf{c}$ (at low temperature) modifies the hidden-order phase and replaces it through a cascade of three first-order transitions by a polarized paramagnetic regime above 39~T [\onlinecite{suslov03}]. The polarized magnetic moment reaches 1.5~$\mu_B$/U at 45~T and continues to increase significantly at higher field, \cite{sugiyama99} showing that the polarization is not complete and that heavy quasi-particles still remain. At low temperature, URu$_2$Si$_2$ is known as a compensated metal. When entering the "hidden-order" phase below $T_0$, a sudden change of the Fermi surface properties has been reported with: i) a decrease by a factor 10 of the density of holes/U (Hall effect \cite{Lerdawson89,kasahara07}) and of the density of electrons/U (thermoelectric power and heat capacity \cite{bel04}), ii) the crossing of the Fermi level by a low-energy quasi-particle band at $T_0$ (angle-resolved-photoemission spectroscopy \cite{santander09} - see also Ref. [\onlinecite{yoshida10,kawasaki11}]), and iii) a strong increase of the carrier mobility (Nernst effect \cite{bel04}). No significant change of the Fermi surface has been seen in the pressure-induced antiferromagnetic state  \cite{hassinger10}, while successive modifications of the Fermi surface were observed when a magnetic field is applied along $\mathbf{c}$ [\onlinecite{jo08,shishido09,altarawneh11}], i.e., when substantial magnetic polarization is induced by the magnetic field. The interplay between the hidden order and the Fermi surface is illustrated by their field-induced variations driven by magnetic polarization effects.

We have performed a systematic investigation of the magnetic and electronic properties of high-purity URu$_2$Si$_2$ single crystals in intense magnetic fields $H$ up to 60~T. Resistivity and magnetization measurements have been carried out with $\mathbf{H}$ along the main crystallographic axes $\mathbf{a}$ and $\mathbf{c}$. The magnetic field - temperature ($H$,$T$) phase diagram, for $\mathbf{H} \parallel \mathbf{c}$, was studied for the first time in both extended temperature (up to 80 K) and magnetic field (up to 60 T) scales. It indicates that that the critical area [35~T-39~T] is initiated by the vanishing of a crossover temperature, which reaches 40-50~K at zero-field. It is demonstrated that this crossover, which probably results from inter-site correlations, is a precursor of the hidden-order phase - this constitutes a new step for the future understanding of hidden-order in URu$_2$Si$_2$. For the first time, magnetoresistivity experiments have been performed up to 60~T within a wide range of transverse and longitudinal configurations, for a magnetic field applied along $a$ and $c$, and by comparing samples of different purities. The characterization of the orbital effect in the magnetoresistivity shows that the Fermi surface is modified at $T_0$ and in a high magnetic field applied along $\mathbf{c}$. The possibility of a field-induced change of the Fermi surface inside the hidden-order phase is emphasized. The carrier mobility is enhanced below the "hidden-order" temperature $T_0$ and decreases close to the high-field polarized regime. In particular for the hidden-order phase, $f$-electrons behavior is intimately connected to the properties of the Fermi surface. This underlines the dual localized-itinerant nature of the $5f$ electrons.

\section{Experimental details} \label{section2}

The URu$_2$Si$_2$ single crystals studied here have been grown by the Czochralski method in a tetra-arc furnace. Resistivity measurements were carried out within the four-point technique on three samples : samples $\sharp1$ and $\sharp2$ with $\mathbf{U},\mathbf{I} \parallel \mathbf{a}$ and the transverse configurations ($\mathbf{H} \parallel \mathbf{c}$; $\mathbf{U},\mathbf{I} \perp \mathbf{H}$) and ($\mathbf{H} \parallel \mathbf{a}$; $\mathbf{U},\mathbf{I} \perp \mathbf{H}$), and sample $\sharp3$ with $\mathbf{U},\mathbf{I} \parallel \mathbf{c}$ and the longitudinal configuration ($\mathbf{H} \parallel \mathbf{c}$; $\mathbf{U},\mathbf{I} \parallel \mathbf{H}$), where $U$ and $I$ are the voltage and current, $H$ is the magnetic field, and $a$ and $c$ are the hard and easy magnetic axes, respectively. Samples $\sharp1$ and $\sharp2$ had residual resistivity ratios $RRR=\rho_{x,x}\rm{(300~K)}$$/\rho_{x,x}\rm{(2~K)}=$~90 and 225, respectively (cf. [\onlinecite{matsuda11}] for a careful investigation of the sample-dependence in the electronic properties of URu$_2$Si$_2$ single crystals), while sample $\sharp3$ had a residual resistivity ratio $RRR=\rho_{z,z}\rm{(300~K)}$$/\rho_{z,z}\rm{(2~K)}=$~85. Zero-field resistivity was measured using the lock-in technique with excitation frequencies of about 17-200 Hz. High-field magnetoresistivity was measured using a digital lock-in (developed at the LNCMI by E. Haanappel) with excitation frequencies of about 20-60 kHz. High-field magnetization was measured using the compensated-coils technique. To estimate the thermal gradients inherent to the compensated-coils set-up, additional torque experiments (which also probe the magnetization, but with less thermal gradients) were performed at temperatures below 8~K. For these experiments, pulsed magnetic fields up to 60 T were generated by standard 6-mm and 20-mm inner bore magnets, with duration times of 150 and 300 ms, respectively, at the high-field facility of the Laboratoire National des Champs Magn\'{e}tiques Intenses at Toulouse.

\begin{figure}[b]
    \centering
    \epsfig{file=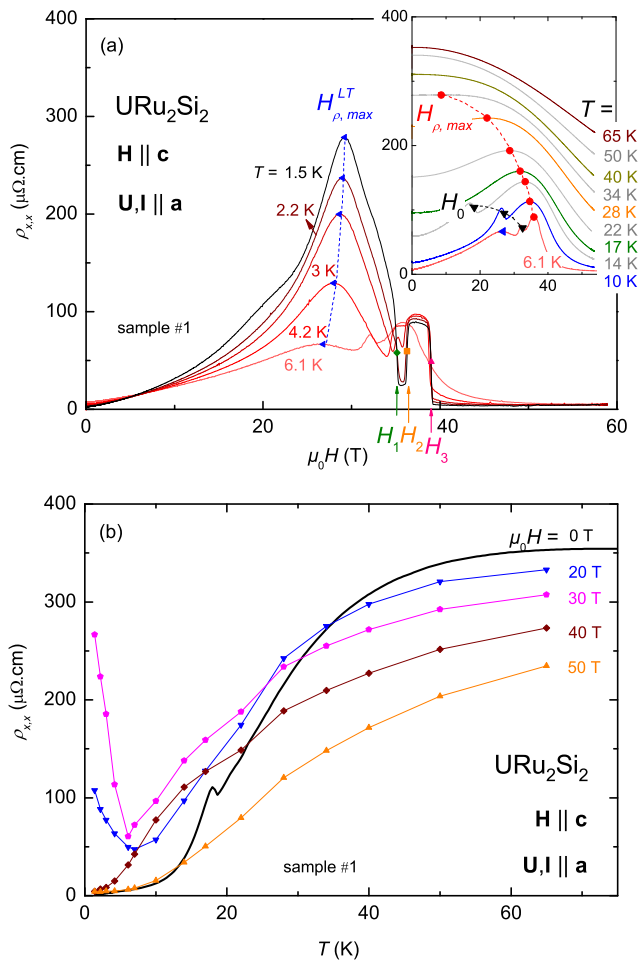,width=85mm}
    \caption{(Color online) (a) Magnetoresistivity $\rho_{x,x}$ versus the magnetic field $H$ (applied along $c$) of sample $\sharp1$ at temperatures between 1.5~K and 6.1~K (between 6.1~K and 65~K in the Inset). (b) Magnetoresistivity versus temperature of sample $\sharp1$ in the magnetic fields $\mu_0H=0,20,30,40$ and 50~T applied along $c$.}
    \label{fig1}
\end{figure}

\section{High-magnetic field properties - (H,T) phase diagram} \label{section3}

\begin{figure}[b]
    \centering
    \epsfig{file=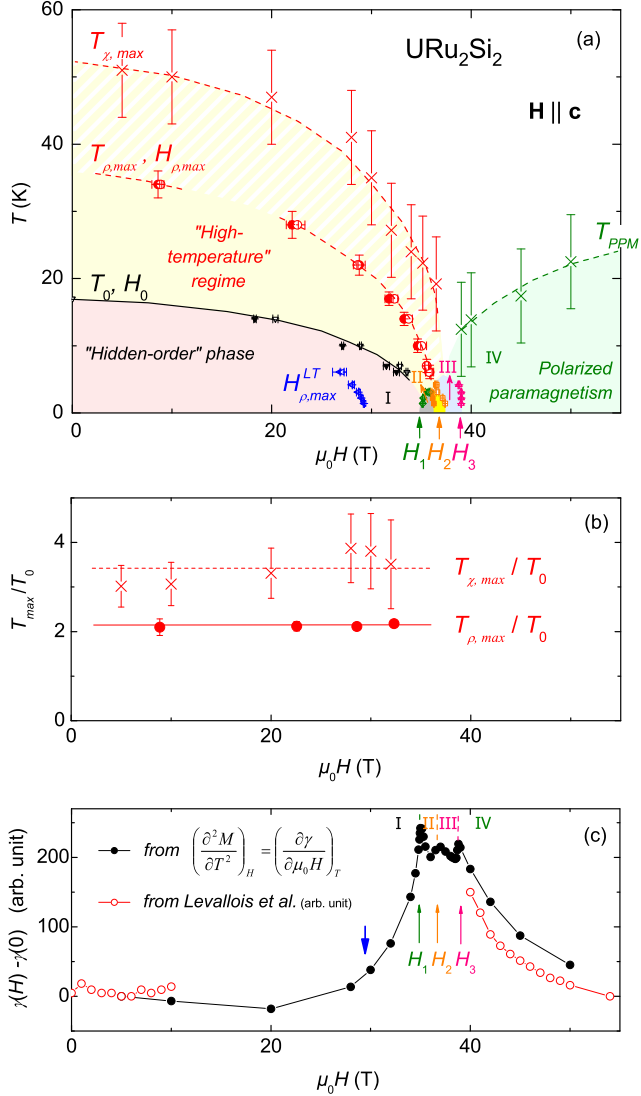,width=85mm}
    \caption{(Color online) (a) Magnetic field-temperature phase diagram of URu$_2$Si$_2$ constructed from our high-field resistivity and magnetization experiments for $\mathbf{H}\parallel\mathbf{c}$. (b) Magnetic field-dependence of the ratios $T_{\rho,max}/T_0$ and $T_{\chi,max}/T_0$, for $\mathbf{H}\parallel\mathbf{c}$. (c) Comparison of the field-dependence of the Sommerfeld coefficient extracted from our magnetization data and that extracted by Levallois \textit{et al.} \cite{levallois09} from resistivity experiments for $\mathbf{H}\parallel\mathbf{c}$. The blue arrow corresponds to the crossover field $H^{LT}_{\rho,max}$ observed in the magnetoresistivity.}
    \label{fig3}
\end{figure}

\begin{figure}[t]
    \centering
    \epsfig{file=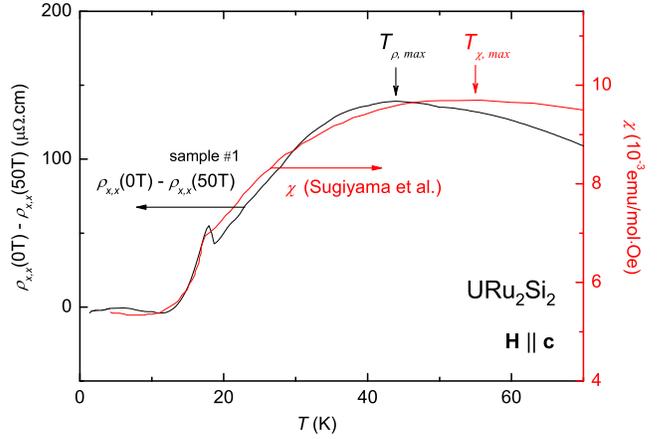,width=85mm}
    \caption{(Color online) Comparison of $\rho_{x,x}(T,\rm{0T})-\rho_{x,x}$$(T,\rm{50T})$ and $\chi(T)$ (from Sugiyama \textit{et al.} \cite{sugiyama99}) versus temperature.}
    \label{fig2a}
\end{figure}

\begin{figure}[b]
    \centering
    \epsfig{file=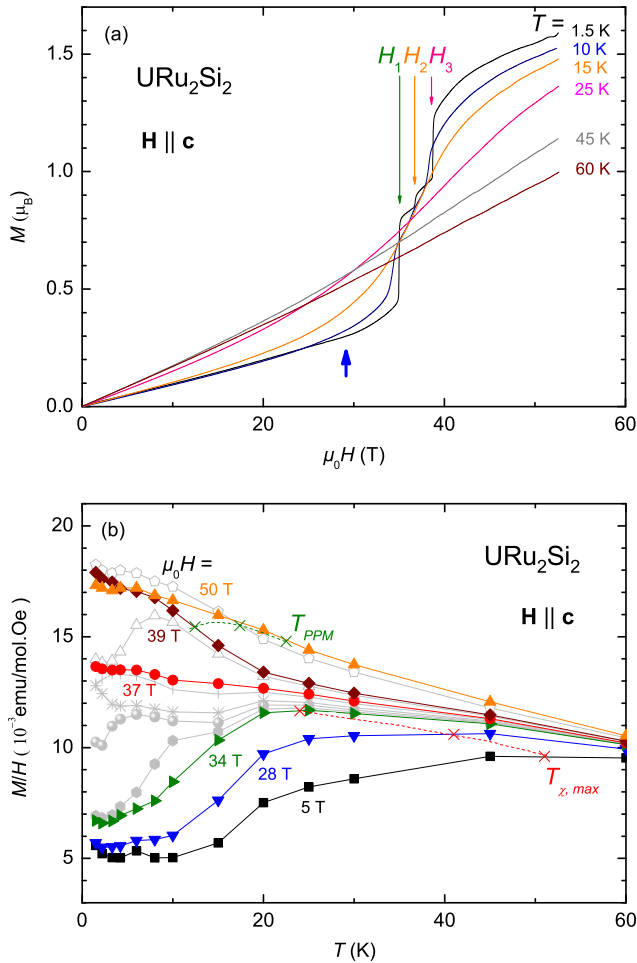,width=85mm}
    \caption{(Color online) (a) Magnetization $M$ versus the magnetic field $H$ (applied along $c$) of URu$_2$Si$_2$ at temperatures between 1.5~K and 60~K. The blue arrow corresponds to the crossover field  $H^{LT}_{\rho,max}$ observed in the magnetoresistivity. (b) Magnetization divided by the magnetic field $M/\mu_0H$ versus temperature at various magnetic fields $\mathbf{H}\parallel \mathbf{c}$ (of 5, 28, 34, 34.5, 35, 35.5, 36.5, 37, 38.5, 39, 45, and 50~T).}
    \label{fig2}
\end{figure}

Figure \ref{fig1} presents transverse resistivity measurements performed on URu$_2$Si$_2$ (sample $\sharp1$) in a high magnetic field $\mathbf{H}$ up to 60~T applied along the easy axis $\mathbf{c}$, at temperatures from 1.5~K to 65~K. Several anomalies are characteristic of magnetic phase transitions at the magnetic fields $H_0$, $H_1$, $H_2$, and $H_3$, and magnetic crossovers at the magnetic fields $H_{\rho,max}$ and $H^{LT}_{\rho,max}$. These transition and crossover lines are described in detail below. They are reported in the phase diagram of Figure \ref{fig3} (a), which agrees well with the phase diagrams established in smaller field and/or temperature ranges (up to 16~K) in Ref. [\onlinecite{suslov03,jaime02,kim03,kim04}]. Below 5~K, three transitions are observed at $\mu_0H_1=35.1\pm0.1$~K, $\mu_0H_2=37.4\pm0.1$~T for increasing field and $36.3\pm0.1$~T for decreasing field, and $\mu_0H_3=39.0\pm0.1$~T, which are defined at the extrema of $\partial\rho_{x,x}/\partial H$. The state below $H_1$ is labeled I and corresponds to the "low-field" hidden-order paramagnetic state. The state labeled II between $H_1$ and $H_2$ and the state labeled III between $H_2$ and $H_3$ correspond presumably to high-field-induced canted antiferromagnetic structures, \cite{sugiyama90} and the state labeled IV above $H_3$ corresponds to the high-field polarized paramagnetic regime. Below 17.5~K, a transition at the magnetic field $H_0$, defined at the extremum of $\partial\rho_{x,x}/\partial H$, corresponds to the boundary of the hidden-order phase. We note equivalently by $H_0(T)$ or $T_0(H)$ this transition line. Although they both delimitate the hidden-order phase, $H_0$ and $H_1$ are two different transition lines, since they lead to different higher-field states. Below 40~K, a "high-temperature" maximum of $\rho_{x,x}$ is observed at $H_{\rho,max}$. The decrease with $T$ of $H_{\rho,max}$ is equivalent in the ($H$,$T$) phase diagram to the decrease with $H$ of a high-temperature crossover scale $T_{\rho,max}$, which reaches 40~K at zero-field and vanishes in the critical field area [35~T-39~T]. However, a maximum in the zero-field resistivity is observed in the bare data at 70~K, but not at 40~K. This can be explained by the fact that an electron-phonon scattering contribution $\rho^{e-ph}_{x,x}$ adds to the purely electronic term $\rho^{e-e}_{x,x}$. A difficulty is to estimate $\rho^{e-ph}_{x,x}(T)$. If we assume that, at 50 T, the magnetic polarization is accompanied by a quenching of almost all magnetic fluctuations and by a vanishing of $\rho^{e-e}_{x,x}$, we can approximate $\rho^{e-ph}_{x,x}$ by $\rho_{x,x}(\rm{50T})$. Following this, we can estimate the purely electronic term by $\rho^{e-e}_{x,x}(T,\rm{0T})$$=\rho_{x,x}(T,\rm{0T})$$-\rho_{x,x}(T,\rm{50T})$ (cf. Fig. \ref{fig2a}). The shift between the maximum observed in $\rho^{e-e}_{x,x}(T,\rm{0T})$ at 40~K and that observed in $\rho_{x,x}(T,\rm{0T})$ at 70~K is due to the additional electron-phonon contribution to the resistivity. The temperature scale of 40 K found in our estimation of $\rho^{e-e}_{x,x}(T,\rm{0T})$ corresponds to $T_{\rho,max}$ extracted from our $\rho_{x,x}(H)$ data, indicating that they correspond to the same phenomenon. Fig. \ref{fig2a} shows a striking similarity between the general shape of $\rho^{e-e}_{x,x}(T,\rm{0T})$ and that of the magnetic susceptibility $\chi(T)$ (from Ref. [\onlinecite{sugiyama90}]). The maxima of $\rho^{e-e}_{x,x}(T,\rm{0T})$ at $T_{\rho,max}\simeq40$~K and of $\chi(T)$ at $T_{\chi,max}\simeq55$~K are thus presumably related to the same physical phenomenon, i.e. a crossover frontier between a high-temperature independent-U-ions regime and a low-temperature interacting-U-ions regime subject to intersite electronic correlations. Below 6~K, a "low-temperature" maximum of $\rho_{x,x}$ is observed at $H^{LT}_{\rho,max}$. $H^{LT}_{\rho,max}$ reaches $29.3\pm0.05$~T at 1.5~K and decreases with increasing $T$, its trace being lost above 6.1~K, where it equals $26.8\pm0.5$~T. The maximum of $\rho_{x,x}$ at $H^{LT}_{\rho,max}$ is associated to orbital effects, i.e., the field-induced motion of quasiparticles along their Fermi surface trajectories, and is a signature of a modification on the Fermi surface (see below). $H^{LT}_{\rho,max}$ coincides with the field above which $M(H)$ becomes non-linear at low temperature, due an enhancement of the magnetic fluctuations (cf. the magnetization in Figure \ref{fig2} and the extracted Sommerfeld coefficient in Figure \ref{fig3} (c)). This crossover at $H^{LT}_{\rho,max}$ is not a phase transition, and it occurs inside the hidden-order phase. Figure \ref{fig1} (b) presents $\rho_{x,x}$ versus $T$ at different magnetic fields, emphasizing that the maximum of $\rho_{x,x}$ at $H^{LT}_{\rho,max}\simeq 30$~T suddenly develops below 6~K, but also that $\rho_{x,x}$ is strongly magnetic field-dependent, for $\mathbf{H}\parallel\mathbf{c}$, at high-temperature (at least up to 65~K).

Figure \ref{fig2} (a) presents the magnetization $M$ versus $H$ at different temperatures $1.5<T<60$~K. In agreement with previous studies, \cite{sugiyama90,sugiyama99} clear anomalies are only observed in $M(H)$ at $H_1$, $H_2$, and $H_3$ which correspond to a succession of first-order transitions between 35 and 39~T leading to a polarized paramagnetic regime above 39~T. $M(H)$ reaches 1.5~$\mu_B$/U at 45~T and continues to increase significantly at higher field, showing that the polarization is not complete due to remaining unquenched magnetic fluctuations. We note that the destruction of the hidden-order state at $H_0$ gives rise to a clear anomaly in  $\rho_{x,x}(H)$ (see Figure \ref{fig1}), but not in $M(H)$. Figure \ref{fig2} (b) shows $M/H$ versus $T$ at different magnetic fields, indicating that a change of behavior occurs in the "cascade" regime [35~T-39~T]. For $\mu_0H<35$~T, $M/H$ is characterized by a broad maximum at the temperature $T_{\chi,max}$ while, for $\mu_0H>39$~T, $M/H$ decreases monotonically with $T$. $T_{\chi,max}$ equals $\simeq50$~K at 5~T and decreases with $H$ before vanishing above 35~T. Between 35~T and 39~T, the cascade of low-temperature transitions $H_1$, $H_2$, and $H_3$ leads to complex features in the $M/H$ versus $T$ plots. Above 39~T, the system becomes polarized paramagnetically, having then a strong field-induced magnetization. The characteristic temperature $T_{PPM}$ of the polarized regime at a given field can be defined at the onset of the enhanced magnetization, i.e., at the inflection point of the $M/H$ versus $T$ curve. $T_{\chi,max}$ and $T_{PPM}$ are reported in the phase diagram shown in Figure \ref{fig3} (a). From similar data than ours but plotted as $\partial M/\partial H$ versus $H$, Sugiyama \textit{et al.} \cite{sugiyama99} have drawn a phase diagram with an almost temperature-independent anomaly, observed up to 60~K, at 40~T. Drawing a phase diagram from $M/H$ versus $T$ plots permitted us to extract the temperature scales $T_{\chi,max}$ of the low-field regime controlled by intersite correlations and $T_{PPM}$ of the high-field polarized regime.

The high-temperature crossover line (denoted by $T_{\rho,max}$ or $H_{\rho,max}$) extracted from our resistivity measurements and the crossover line $T_{\chi,max}$ defined in our magnetization data are surely controlled by the same phenomenon (as shown by the constance of the ratio $T_{\rho,max}/T_{\chi,max}$ but also by the similar temperature-dependences of $\rho^{e-e}_{x,x}(T,\rm{0T})$ and $\chi(T)$ in Fig. \ref{fig2a}). The offset between $T_{\rho,max}$ and $T_{\chi,max}$ in Figure \ref{fig3} (a) is due to the difficulty to define precisely the temperature or magnetic field of a crossover, that is, to the non-equivalence of their definitions. As shown in Figure \ref{fig3} (b), the ratios $T_{\rho,max}/T_0$ and $T_{\chi,max}/T_0$ are both constant up to 35 T. This indicates that the vanishing of the higher-temperature crossover scale (either $T_{\rho,max}$ or $T_{\chi,max}$) controls that of $T_0$. In other words, the mechanism responsible for the crossover at $T_{\rho,max}$ or $T_{\chi,max}$ is a precursor of the hidden-order state since its destabilization leads to that of the hidden order, and the high-temperature regime is a necessary condition for the development of the hidden-order state. The field-induced vanishing of the high-temperature crossover, which is the mark of intersite electronic correlations,  governs thus both the critical area leading to a polarized regime above 39 T and the destabilization of the hidden-order state at low temperature.

Figure \ref{fig3} (c) presents the field-dependence of the Sommerfeld coefficient $\gamma\underset{T\rightarrow0}{=}C_p/T$, where $C_p$ is the specific heat, estimated using the Maxwell relation:
\begin{eqnarray}
\left(\frac{\partial\gamma}{\partial \mu_0H}\right)_T=\left(\frac{\partial^2M}{\partial T^2}\right)_H,
    \label{sommerfeld}
\end{eqnarray}
and assuming that $M(T,H)=M(0,H)-\beta T^2$ is obeyed (cf. also Ref. [\onlinecite{paulsen90}]). Because of thermal gradients in our pulsed-field magnetization probe (for the fits, the temperature was corrected thanks to additional torque experiments), the variation of $\gamma$ extracted here is only qualitative and expressed in arbitrary units. A strong enhancement of $\gamma$, and thus of the effective mass $m^*$, is found in a broad magnetic field window between 30~T and 45~T. A comparison with the $H$-variation of $\sqrt A$, which also probes $m^*$ assuming the validity of a Fermi liquid picture (where $A$ is the quadratic coefficient of the resistivity from [\onlinecite{levallois09})] in a frame where magnetic fluctuations dominates, shows a qualitative agreement between the two methods.

\section{"Orbital" magnetoresistivity - Fermi surface reconstructions} \label{section4}

\begin{figure}[t]
    \centering
    \epsfig{file=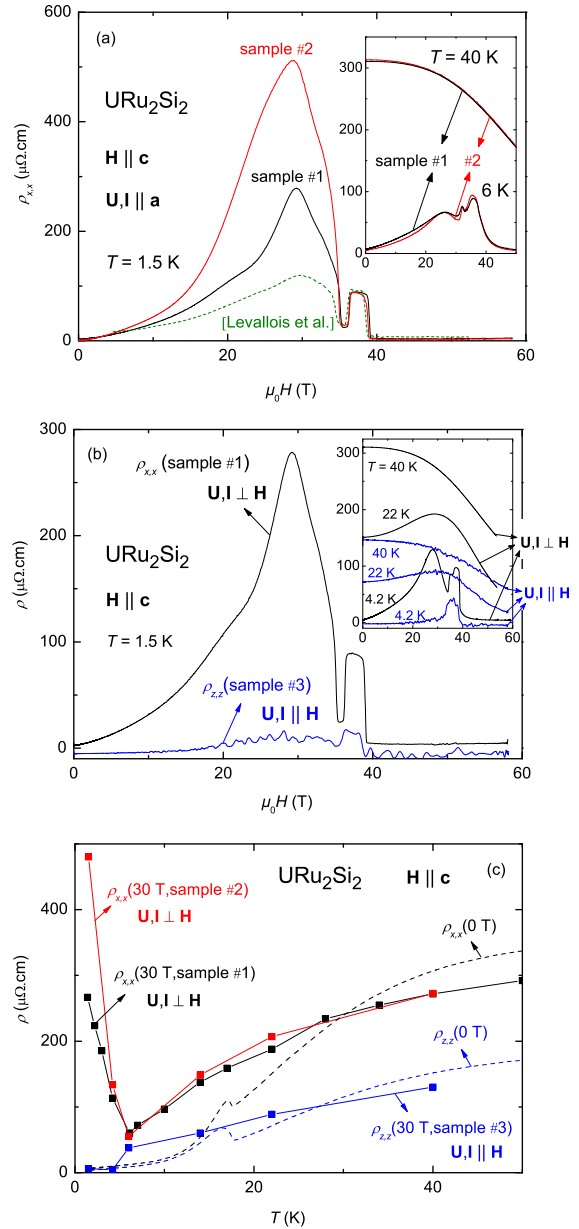,width=75mm}
    \caption{(Color online) (a) Transverse magnetoresistivity $\rho_{x,x}$ versus magnetic field, for $\mathbf{H}\parallel\mathbf{c}$, at $T=1.5$~K measured here on samples $\sharp1$ and $\sharp2$ and by Levallois \textit{et al.} on a third sample \cite{levallois09}. The Inset shows $\rho_{x,x}(H)$ at $T=6$ and 40~K for samples $\sharp1$ and $\sharp2$. (b) Comparison of the transverse and longitudinal magnetoresistivity $\rho_{x,x}$ and $\rho_{z,z}$ versus $H$ of samples $\sharp1$ and $\sharp3$, respectively, for $\mathbf{H}\parallel\mathbf{c}$ at $T=1.5$~K. The Inset shows $\rho_{x,x}(H)$ and $\rho_{z,z}(H) $at $T=4.2$, 22 and 40~K. (c) Temperature-dependence of $\rho_{x,x}$ of samples $\sharp1$ and $\sharp2$ and $\rho_{z,z}$ of sample $\sharp3$, at $\mu_0H=0$ and 30~T, for $\mathbf{H}\parallel\mathbf{c}$.}
    \label{fig4}
\end{figure}

Figure \ref{fig4} (a) shows a plot of the transverse magnetoresistivity $\rho_{x,x}$ measured for $\mathbf{H} \parallel \mathbf{c}$ and $\mathbf{U},\mathbf{I} \parallel \mathbf{a}$ on two samples noted $\sharp1$ and $\sharp2$ studied here, and a third sample studied by Levallois \textit{et al.} \cite{levallois09} $\rho_{x,x}$ is almost sample-independent above 35~T, that is, in the regime controlled by the cascade of magnetic phase transitions (similar values of $\rho_{x,x}$ above 35~T have been found in Ref. [\onlinecite{jo07,oh07}]) and is characterized by a strongly sample-dependent maximum at 30~T. $\rho_{x,x}$ at the top of this anomaly is twice bigger for sample $\sharp2$ than for sample $\sharp1$, where it is three times bigger than for the sample studied by Levallois \textit{et al.} \cite{levallois09} Knowing that samples $\sharp1$, $\sharp2$, and that studied by Levallois \textit{et al.} \cite{levallois09} have residual resistivity ratios $\rho_{x,x}\rm{(300~K)}$$/\rho_{x,x}\rm{(2~K)}\simeq$~90, 225, and 40, respectively, we find that the higher the quality of the sample is, the bigger is the anomaly in $\rho_{x,x}$ at 30~T. This is compatible with the strong magnetoresistivity reported by Kasahara \textit{et al.} \cite{kasahara07} at low temperature and up to 10~T on high-quality single crystals of URu$_2$Si$_2$. We note that a change of curvature is observed at $T=1.5$~K in $\rho_{x,x}(H)$ at around 20-25~T for sample $\sharp1$ (and for the sample studied by Levallois \textit{et al.} \cite{levallois09}) but not for our best sample (sample $\sharp2$). Shishido \textit{et al.} \cite{shishido09} reported a similar anomaly in $\rho_{x,x}$ below 1~K (in a crystal of similar quality than sample $\sharp2$) and interpreted it as a transition driven by a Fermi surface reconstruction. When the temperature is increased, as shown at $T=6$~K and 40~K in the inset of Figure \ref{fig4} (a), the magnetoresistivity becomes almost sample-independent.

In figures \ref{fig4} (b-c), a comparison is made between the transverse and longitudinal magnetoresistivities $\rho_{x,x}$ and $\rho_{z,z}$, respectively, measured at 1.5~K on two samples ($\sharp1$ and $\sharp3$) of similar qualities ($\rho_{z,z}\rm{(300~K)}$$/\rho_{z,z}\rm{(2~K)}=$~85 for sample $\sharp3$) in a field $\mathbf{H}\parallel\mathbf{c}$. In spite of a bigger noise (due to the smaller resistance of sample $\sharp3$), $\rho_{z,z}$ presents similar anomalies than $\rho_{x,x}$ at $H_{\rho,max}$, $H_0$, $H_1$, $H_2$, and $H_3$. At high temperature, the difference between the absolute values of $\rho_{x,x}$ and $\rho_{z,z}$ versus $H$ reflects their different behaviors at zero-field (cf. Inset of Figure \ref{fig4} (b) and Figure \ref{fig4} (c)). At low temperature, this contribution to the magnetoresistivity leads to a maximum at 30~T in the transverse configuration, but not in the longitudinal configuration (Figure \ref{fig4} (b)). This is confirmed in Figure \ref{fig4} (c), where a sudden increase of $\rho_{x,x}$ (samples $\sharp1$ and $\sharp2$) occurs below 6~K at $\mu_0H=30$~T, while the longitudinal resistivity $\rho_{z,z}$ (sample $\sharp3$) at $\mu_0H=30$~T decreases below 6~K. In this last configuration, only a few temperatures have been investigated, and the 'apparent' sudden decrease below 6~K of $\rho_{z,z}$ measured at 30~T is related to the anomaly at $T_0$ (which is reduced at 30~T compared to $T_0$ at zero-field) but not to an orbital effect as in the transverse configuration.

To be concise, the maximum of magnetoresistivity observed in a magnetic field of 30~T applied along $\mathbf{c}$ i) develops at low temperature (below 6~K), ii) is present in the transverse configuration, but not in the longitudinal configuration, and iii) is enhanced when the sample quality, and thus the electronic mean free path, are higher. We can safely conclude that this anomaly is controlled by a field-induced cyclotron motion of the conduction electrons, that is, an orbital contribution to the magnetoresistivity within the condition $\omega_c\tau>1$, where $\omega_c$ is the cyclotron frequency and $\tau$ is the lifetime of the conduction electrons. A modification of the Fermi surface accompanied by a reduction of the carrier mobility $\mu=\omega_c\tau/\mu_0H$ is a natural way to explain the decrease of $\rho_{x,x}$ above 30~T. It is worthwhile to remark that an enhancement of critical magnetic fluctuations, as indicated by the field-dependence of the Sommerfeld coefficient is also observed  above 30~T (Figure \ref{fig3} (c)). This underlines the strong interplay between the magnetic polarization and the field-induced evolution of the Fermi surface in URu$_2$Si$_2$. In fact, recent Shubnikov-de Haas experiments clearly demonstrated that a new frequency emerges above $\sim25$~T, while the Fermi surface branch $\alpha$ shrinks in volume \cite{altarawneh11,aoki12}. Oppositely, the anomalies in $\rho_{x,x}$ at $H_{\rho,max}$, $H_0$, $H_1$, $H_2$, and $H_3$, whose shape and size do not change with the sample quality, are independent of $\omega_c\tau$. We can assert that the scattering off of $f$-electrons is sample-independent, since it corresponds to a scattering of conduction electrons by the static or fluctuating magnetic moments from each $5f$ U-ion site, the distance - between two ions - involved in this process being smaller than the distance between two impurities.

\begin{figure}[t]
    \centering
    \epsfig{file=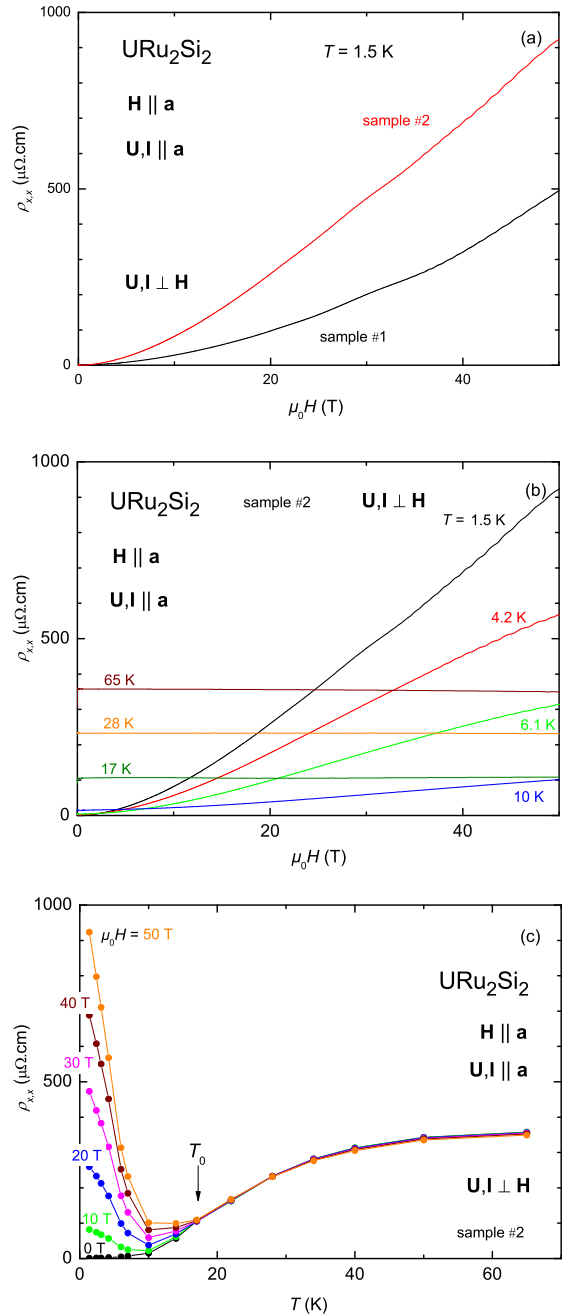,width=74mm}
    \caption{(Color online) (a) Transverse magnetoresistivity $\rho_{x,x}$ versus magnetic field, for $\mathbf{H}\parallel\mathbf{a}$, at $T=1.5$~K measured here on samples $\sharp1$ and $\sharp2$. (b) Transverse magnetoresistivity $\rho_{x,x}$ versus magnetic field, for $\mathbf{H}\parallel\mathbf{a}$ and at different temperature between 1.5 and 65~K, of sample $\sharp2$.(c) Temperature-dependence of $\rho_{x,x}$ of sample $\sharp2$, at the constant fields $\mu_0H=0$, 10, 20, 30, 40 and 50~T, for $\mathbf{H}\parallel\mathbf{a}$.}
    \label{fig5}
\end{figure}

Figure \ref{fig5} presents the magnetoresistivity $\rho_{x,x}$ of samples $\sharp1$ and $\sharp2$ in a magnetic field parallel to the hard magnetic axis $\mathbf{a}$, within a transverse configuration $\mathbf{U},\mathbf{I}\perp\mathbf{H}$ [\onlinecite{note}]. In Figure \ref{fig5} (a), the magnetoresistivity at 1.5~K increases monotonically with the magnetic field, being almost a factor two bigger in sample $\sharp2$ than in sample $\sharp1$ due to a higher mean free path in sample  $\sharp2$. Quantum oscillations are also observed for the two compounds and confirm their high quality (a forthcoming paper \cite{scheerer12} will focus on their analysis). Figures \ref{fig5} (b) and (c) show that the strong magnetic field-dependence of $\rho_{x,x}$ under $\mathbf{H}\parallel\mathbf{a}$ is reduced when $T$ is increased, which is the signature of an impurity-dependent signal of orbital origin, as well as the 30-T-anomaly in $\rho_{x,x}$ under $\mathbf{H}\parallel\mathbf{c}$. A striking feature is the sudden suppression of the magnetic field-dependence of $\rho_{x,x}$ above the hidden-order transition temperature $T_0=17.5$ K when $\mathbf{H}\parallel\mathbf{a}$. This result is compatible with a Fermi surface reconstruction occurring at $T_0$, with a strong reduction of the electronic mobility above $T_0$, as suggested in Ref. [\onlinecite{Lerdawson89,kasahara07,bel04,santander09,yoshida10}]. For $\mathbf{H}\parallel\mathbf{a}$ (Fig. \ref{fig5} (c)), a sudden change of the $H$-dependence of $\rho_{x,x}$, due to a modification of the orbital term, is easily observed below $T_0$ since there is no additional variation with $H$ of $\rho_{x,x}$ driven by field-induced magnetic properties. For $\mathbf{H}\parallel\mathbf{c}$ (Fig. \ref{fig1} (b)), the situation is different: there is a significant variation with $H$ of $\rho_{x,x}$  at all temperatures (up to 65~K here) due to a field-induced modification of the magnetic properties. This magnetic contribution adds to the orbital contribution to $\rho_{x,x}$, and we cannot determine precisely the temperature below which the orbital contribution develops. At high-temperature ($T>T_{\chi,max}$), the magnetic field quenches the scattering of conduction electrons on $f$-electrons moments through a negative slope of the magnetoresistivity versus field, with an amplitude which depends on the size of the magnetization. This effect is strong for $\mathbf{H}$ along the easy magnetization axis $\mathbf{c}$ and very small for $\mathbf{H}$ along the hard axis $\mathbf{a}$.

\section{Discussion} \label{section5}

In URu$_2$Si$_2$, the magnetic-field dependences of $T_{\chi,max}$ and $T_{PPM}$ (Figure \ref{fig3} (a)), as well as the plots of $M/H$ versus $T$ (Figure \ref{fig2} (b)), recall strongly the case of the heavy-fermion paramagnet CeRu$_2$Si$_2$  [\onlinecite{paulsen90,ishida98}], which is characterized by a pseudo-metamagnetic transition to a polarized state at $H_m=7.8$~T. As already shown in Ref. [\onlinecite{inoue11}], a correspondence 1~K~$\leftrightarrow$~1~T relates the maximum of susceptibility $T_{\chi,max}$ to the critical magnetic field $H^*$ of several heavy-fermion systems (including URu$_2$Si$_2$, for which $H^*=35-39$~T, and CeRu$_2$Si$_2$, for which $H^*=H_m=7.8$~T cf. Table \ref{table}), suggesting that both $T_{\chi,max}$ and $H^*$ are controlled by a single magnetic energy scale. In CeRu$_2$Si$_2$, $T_{\chi,max}$ and $H_m$ are controlled by antiferromagnetic fluctuations at the wavevector $\mathbf{k}_1=(0.31,0,0)$, since i) antiferromagnetic fluctuations at $\mathbf{k}_1$ vanish at $H_m$ [\onlinecite{flouquet04}] and ii) their energy scale, that is, the low-temperature quasielastic linewidth $\Gamma_1=10$~K [\onlinecite{raymond07,knafo09}], is also equal to $T_{\chi,max}$ [\onlinecite{fisher91}] (cf. Table \ref{table}). The regime below $T_{\chi,max}$ in URu$_2$Si$_2$ is probably related to the onset of intersite magnetic correlations too. Correlations in URu$_2$Si$_2$ might be more complex than in CeRu$_2$Si$_2$ due to the possible interplay between multipolar interactions and the crystal field \cite{haule09}. However, it is known that above $T_0$ antiferromagnetic fluctuations at $\mathbf{Q}_1$ have a linewidth $\Gamma_1\simeq50$~K~$\simeq T_{\chi,max}$  (with a gap of $\sim25$~K), \cite{broholm91,palstra85} which recalls the CeRu$_2$Si$_2$ case. It is suspected that the large damping at high temperature prevents the detection of the low-energy gap $\Delta_0$ at $\mathbf{Q}_0$, which appears below $T_0$ as a sharp resonance. Below $T_0$, the case of URu$_2$Si$_2$ is undoubtedly more complex than the CeRu$_2$Si$_2$ case, since the linewidth at $\mathbf{Q}_1$ is significantly reduced, being accompanied by a strong modification of the electronic density of states. Extrapolation of neutron data measured up to 20~T [\onlinecite{bourdarot03}] indicates that, at low-temperature, the destruction of hidden-order at 35 T might occur when the gaps $\Delta_1$ and $\Delta_0$ in the excitation spectra at $\mathbf{Q}_1$ and $\mathbf{Q}_0$, respectively, converge to a common value. This is compatible with a loss of dispersion, that is, with a loss of wavevector-dependent antiferromagnetic correlations above 35~T. Alternatively, $T_{\chi,max}$ in URu$_2$Si$_2$ could be controlled by intersite correlations between high-order multipoles, such as the hexadecapoles (instead of the dipole moments) which, in the model proposed by Kusunose and Harima \cite{kusunose11}, would order in the hidden-order state. For both URu$_2$Si$_2$ and CeRu$_2$Si$_2$, a change from intersite multipolar (bipolar or of a higher-order) interactions to a high-field polarized paramagnetic regime occurs when the magnetic polarization reaches 0.5~$\mu_B/$ion. Sweeping a magnetic field along $\mathbf{c}$ initiates a ferromagnetic coupling which becomes dominant when the pseudo-gap built by the intersite correlations is closed, which coincides with the collapse of $T_{\chi,max}$. As in CeRu$_2$Si$_2$ [\onlinecite{flouquet04}], the field-induced enhancement of $m^*$ in URu$_2$Si$_2$ might be related to critical ferromagnetic fluctuations, a switch occurring from a small Fermi surface in the hidden-order state to a large Fermi surface in the polarized regime. Instead of a well-defined quantum critical point, there are several indications for a quantum critical area between 35 and 39~T in URu$_2$Si$_2$: i) the field-temperature phase diagram is made of a low-field regime, whose characteristic temperature $T_{\chi,max}$ vanishes at 35~T, and of a polarized paramagnetic regime above 39~T, ii) the effective mass is enhanced in a wide regime between 35 and 40~T, indicating enhanced and thus critical magnetic fluctuations, and iii) the susceptibility at 37 T looks similar to that of usual quantum critical systems. A singularity of URu$_2$Si$_2$ is that, instead of a unique second-order phase transition at a given critical field, its low-temperature phase diagram is made of a cascade between 35 and 39~T of three first-order transitions at $H_1$, $H_2$, and $H_3$, with an additional sharp crossover at $H^{LT}_{\rho,max}\simeq30$~T within the hidden-order phase.

\begin{table}[tb]
\caption{Critical magnetic field, maximum of the magnetic susceptibility, and energy scale of the antiferromagnetic correlations in CeRu$_2$Si$_2$ and URu$_2$Si$_2$ [\onlinecite{broholm91,flouquet04,raymond07,knafo09,fisher91,knafo04,matsumoto08,palstra85}].}
\begin{ruledtabular}
\begin{tabular}{lcccc}
&$H^*$&$T_{\chi,max}$&Antiferromagnetic linewidth&\\
\hline
CeRu$_2$Si$_2$&7.8 T&10~K&$\Gamma(\mathbf{k}_1,T\rightarrow 0)=10$~K\\
URu$_2$Si$_2$&35-39~T&55~K&$\Gamma(\mathbf{Q}_1,T\geq T_0)=50$~K\\
\end{tabular}
\end{ruledtabular}
\label{table}
\end{table}

The observation by optical spectroscopy of a hybridization gap above $T_0$ [\onlinecite{levallois12}] might be directly linked to the development of intersite correlations below the high-temperature scale $T_{\chi,max}$ or  $T_{\rho,max}$ determined here. Despite an anomalous Fermi-liquid behavior reported above $T_0$ by spectroscopy in [\onlinecite{nagel12}], the situation above $T_0$ is already well-understood by macroscopic experiments. Without the establishment of hidden-order, the specific heat divided by the temperature $C_p/T$ in URu$_2$Si$_2$ is expected to behave similarly than in usual intermediate-valent systems, such as CeSn$_3$, with a broad maximum at around 30~K [\onlinecite{flouquet11}]. Derivation of the Gr\"{u}neisen parameter $\Omega_T$ clearly shows that a constant value close to 40 will be achieved only at very low temperature, while $\Omega_T$ reaches values close to 18 at $T_0$ and close to 5 at $T_{\chi,max}$. A true Fermi liquid behavior is a regime controlled by a single energy scale and where the Gr\"{u}neisen parameter is temperature-independent. No Fermi-liquid behavior can thus be achieved above $T_0$, where a competition occurs between the different energy scales of the system, and it is not surprising that forcing a Fermi liquid analysis above $T_0$, as done in [\onlinecite{nagel12}], leads to strong deviations from a Fermi liquid description.

The specificity of URu$_2$Si$_2$ is that the Fermi surface reconstruction at $T_0$ leads to different Fermi surface bands whose characteristic bandwidths are rather low, due to the combined effects of low-carrier densities and high effective masses. Applying a magnetic field permits to decouple the minority and majority spin bands. Considering the complex Fermi surface of URu$_2$Si$_2$, exotic phenomena such as a cascade of Lifshitz transitions may occur. At very low temperature ($T\simeq50$~mK), marks of changes in the magnetoresistivity regime were detected already at 8~T [\onlinecite{hassinger10b}], as well as new de Haas - van Alphen frequencies near 21~T [\onlinecite{shishido09}], in the window 21-25~T [\onlinecite{aoki12}], and in the field ranges 17-24~T, 24-29.4~T, and 29-34.7~T [\onlinecite{altarawneh11}]. Deep extrema in the thermoelectric power were also detected at 10 and 24~T [\onlinecite{malone11}]. We note that, up to 30~T, these changes occur in a linear-field magnetization regime where no sign of phase transition has also been reported in the specific heat yet.

As for Ce-based heavy-fermion compounds, \cite{flouquet05} the temperature- and magnetic field-properties of URu$_2$Si$_2$ are governed by the interplay of several electronic energy scales. This interplay is illustrated by a strong temperature-dependence of the electronic Gr\"{u}neisen parameter, which often reaches a constant value only at very low temperature, corresponding then to the entrance into a Fermi liquid state. The key ingredients are the Kondo temperature $T_K$ associated with the renormalized bandwidth, the intersite coupling $E_{i,j}$ between $f$-electron moments, and the crystal-field energy $\Delta_{CF}$. In many heavy-fermion systems, $k_BT_K$, $E_{i,j}$, and $\Delta_{CF}$ are comparable, leading to mixed spin and valence fluctuations. CeCu$_2$Si$_2$ and CeRu$_2$Si$_2$ are two cases where spin and valence fluctuations are well-decoupled, since the hierarchy $\Delta_{CF}>k_BT_K>E_{i,j}$ is well-defined and their quantum instability occurs for a well-defined doublet ground state. \cite{flouquet05} Exotic situations can occur as in Ce$_{1-x}$La$_x$B$_6$, where the crystal-field ground-state is a quartet leading to octupole order. \cite{kuwahara07,shiina07} The interplay between the Kondo effect and intersite interactions generally leads to a pseudo-gap structure in the density of states. By comparison with Ce-based heavy-fermion systems, the novelties of URu$_2$Si$_2$ are i) that valence fluctuations occur between two configurations $5f^2$ and $5f^3$, which both carry a large momentum at high temperature and ii) that a renormalization to a $5f^2$-like configuration would lead to a singlet ground-state. \cite{kuramoto09} Concerning the valence fluctuations, the situation of URu$_2$Si$_2$ [\onlinecite{denlinger11}] might be comparable to that of TmSe where, despite a valence close to 2.5, long-range ordering occurs and is associated with a metal-insulator transition, the Tm-ions being renormalized to a Tm$^{2+}$ state driving the system to an insulating antiferromagnetic phase. \cite{derr06} Assuming that $\Delta_{CF}$ is comparable to $k_BT_K$, Haule and Kotliar \cite{haule09} find a $5f^2$ ground state favoring multipolar ordering, as in Pr$^{3+}$ $4f^2$ skutterudite systems. \cite{hassinger08,kuramoto09b} In fact, the arrested Kondo model developed in [\onlinecite{haule09}] and the multipolar ordering scenario proposed in [\onlinecite{kusunose11,harima10}] both assume that the ground state is governed by the properties of the $5f^2$ configuration, even if the $5f$ electrons are itinerant. Within these scenarios \cite{haule09,kusunose11,harima10}, a magnetic field would modify the order parameter via the modification of the fundamental ground-state and thus of the possible multipolar couplings. This might be compatible with the deep Fermi surface reconstructions established by quantum oscillations coupled to band structure modeling, \cite{hassinger10,hassinger10b} ARPES, \cite{santander09,yoshida10,denlinger11} and scanning tunneling microscopy. \cite{schmidt10,aynajian10} New core spectroscopy experiments \cite{fujimori12} led to the proposal that the valence of URu$_2$Si$_2$ may be close to 3, in rather good agreement with a theoretical model recently developed by Ikeda \textit{et al.} \cite{ikeda12} which predicts a valence of 2.7. In this last approach the hidden-order is considered to be a dotriacontopole (rank five) and the possibility of different channels for the ground state (whatever is the valence) is the mark of strong local properties of the $5f$ electrons.

The dual nature of the $f$-electrons is at the heart of the heavy-fermion problem, where the hybridization of $f$ and conducting electrons affects both the magnetic properties of the f-electrons (reduced magnetic energy scales, damped magnetic fluctuations, etc.) and the conducting properties of the itinerant bands (renormalization of the Fermi surface, etc.). As a result, the $f$-electrons often have both a localized character (density of $f$-electron per site close to 1, well-defined spin waves and crystal-field levels, magnetic entropy close to $R\rm{ln}2$ etc.) and an itinerant character (contribution to the Fermi surface, Fermi liquid behavior, nesting wavevectors, etc.). The case of URu$_2$Si$_2$ is complex, since both magnetic and Fermi surface properties are strongly modified in the hidden-order phase: i) at $T_0$, modifications of the magnetic fluctuations spectra have been probed by inelastic neutron scattering (34) while a Fermi surface reconstruction has been probed by magnetoresistivity, Hall effect, \cite{Lerdawson89,kasahara07} Nernst effect, \cite{bel04} and ARPES, \cite{santander09,yoshida10} ii) a magnetic crossover at 40-50~K is found here to be a precursor of the hidden-order state, since its vanishing leads to the critical area at 35-39~T and drives to the destruction of the hidden order, and iii) in a magnetic field $\mathbf{H}\parallel\mathbf{c}$, a sharp crossover at 30~T is related to a Fermi surface evolution, which occurs when the effective mass (dressed by field-induced ferromagnetic fluctuations) becomes enhanced. Interplay of the $f$-electrons magnetic properties with that of the Fermi surface has been reported in other U-based compounds, such as the ferromagnetic superconductors UGe$_2$, URhGe, and UCoGe where the switch from the paramagnetic to the ferromagnetic phases induces a strong change of the Fermi surface topology (see Refs. [\onlinecite{flouquet05}] for UGe$_2$, [\onlinecite{yelland11}] for URhGe, and [\onlinecite{malone12,aoki11}] for UCoGe). The novelty in U-based intermetallic compounds is that the $5f$ bands are already quite close to the Fermi level and that they are very flat. \cite{mydosh11} Small changes in the Fermi level are expected to generate drastic changes of the Fermi surface, as observed here for URu$_2$Si$_2$ via the strong modifications at $T_0$ or in a high magnetic field of 30 T applied along $\mathbf{c}$ of the orbital contribution to the magnetoresistivity. An appropriate description of the dual "localized-itinerant" nature of the $f$-electrons in URu$_2$Si$_2$ should enable a global understanding of the interplay between the magnetic and Fermi surface properties, which could be a key to solve the hidden-order problem.

\section{Conclusion}

To conclude, we have performed a systematic investigation per magnetoresistivity and magnetization of high-quality URu$_2$Si$_2$ single crystals in pulsed magnetic fields up to 60~T and temperatures between 1.5 and 65~K. We have drawn the magnetic-field-temperature phase diagram of the system for $\mathbf{H}\parallel\mathbf{c}$, in extended scales going up to 60~T and 60~K. A high-temperature crossover probed by magnetoresistivity at $T_{\rho,max}\simeq40$~K and by magnetization at $T_{\chi,max}\simeq55$~K (at zero-field) is related to the onset of intersite electronic correlations and is found to be a precursor of the "hidden-order" phase (which develops below $T_0=17.5$~K at $H=0$). In a magnetic field applied along $\mathbf{c}$, the vanishing of the crossover temperature $T_{\rho,max}$ or $T_{\chi,max}$ is responsible i) for the critical area developing at [35~T-39~T] and ii) for the destabilization of the hidden-order state, a polarized regime being reached above 39~T. Magnetoresistivity measurements on three high-quality single crystals were performed in magnetic fields applied along the hard axis $\mathbf{a}$ and the easy-axis $\mathbf{c}$, for both transverse and longitudinal configurations. A sample-dependent orbital contribution to the magnetoresistivity confirmed that a Fermi surface reconstruction occurs at the hidden-order temperature $T_0$, but also that the Fermi surface is modified in a field of 30~T applied along $\mathbf{c}$. The interplay between the magnetic properties and that of the Fermi surface has been emphasized, as well as the necessity to use a dual "localized-itinerant" description of the $f$-electrons for a future understanding of the hidden-order in URu$_2$Si$_2$.\\

\section*{Acknowledgments}

We acknowledge J. B\'{e}ard, L. Bendichou, P. Delescluse, T. Domps, J-M Lagarrigue, M. Nardone, J.-P. Nicolin, C. Proust, T. Schiavo, and A. Zitouni for technical support and K. Behnia, F. Bourdarot, F. Hardy, H. Harima,H. Kusunose, J. Levallois, and C. Proust for useful discussions. This work was supported by the French ANR DELICE, by Euromagnet II via the EU under Contract No. RII3-CT-2004-506239, and by the ERC Starting Grant NewHeavyFermion.


\begin{thebibliography}{20}
\bibitem{note} In the tetragonal structure of URu$_2$Si$_2$, there are two equivalent $a$-axes. A transverse configuration $\mathbf{U},\mathbf{I}\perp\mathbf{H}$ with $\mathbf{U},\mathbf{I}\parallel\mathbf{a}$ and $\mathbf{H}\parallel\mathbf{a}$ is thus possible.
\end{thebibliography}
\end{document}